\documentclass{SciPost}

\hypersetup{
    colorlinks,
    linkcolor={red!50!black},
    citecolor={blue!50!black},
    urlcolor={blue!80!black}
}

\usepackage{xspace}
\usepackage{float}
\usepackage[bitstream-charter]{mathdesign}
\DeclareSymbolFont{usualmathcal}{OMS}{cmsy}{m}{n}
\DeclareSymbolFontAlphabet{\mathcal}{usualmathcal}

\fancypagestyle{SPstyle}{
\fancyhf{}
\lhead{\colorbox{scipostblue}{\bf \color{white} ~SciPost Physics Proceedings }}
\rhead{{\bf \color{scipostdeepblue} ~Submission }}

\fancyfoot[C]{\textbf{\thepage}}
}

\begin{document}
\newcommand{\ttbar}{$t\overline{t}$\xspace}
\newcommand{\vtb}{$V_{tb}$\xspace}
\newcommand{\bjet}{$b$-jet\xspace}
\newcommand{\bjets}{$b$-jets\xspace}
\newcommand{\etmiss}{$E_{\text{T}}^{\text{miss}}$\xspace}
\newcommand{\mtw}{$m_{\text{T}}(W)$\xspace}
\newcommand{\flvvtb}{$|f_{LV}V_{tb}|$\xspace}

\pagestyle{SPstyle}

\begin{center}{\Large \textbf{\color{scipostdeepblue}{
Inclusive top cross sections in ATLAS\footnote{Copyright 2022 CERN for the benefit of the ATLAS Collaboration. Reproduction of this article or parts of it is allowed as specified in the CC-BY-4.0 license.}\\
}}}\end{center}

\begin{center}\textbf{
Charlie Chen, on behalf of the ATLAS Collaboration\textsuperscript{1$\star$}
}\end{center}

\begin{center}
{\bf 1} University of Victoria
\\[\baselineskip]
$\star$ \href{mailto:bohan.chen@cern.ch}{\small bohan.chen@cern.ch}
\end{center}

\definecolor{palegray}{gray}{0.95}
\begin{center}
\colorbox{palegray}{
  \begin{tabular}{rr}
  \begin{minipage}{0.36\textwidth}
    \includegraphics[width=60mm,height=1.5cm]{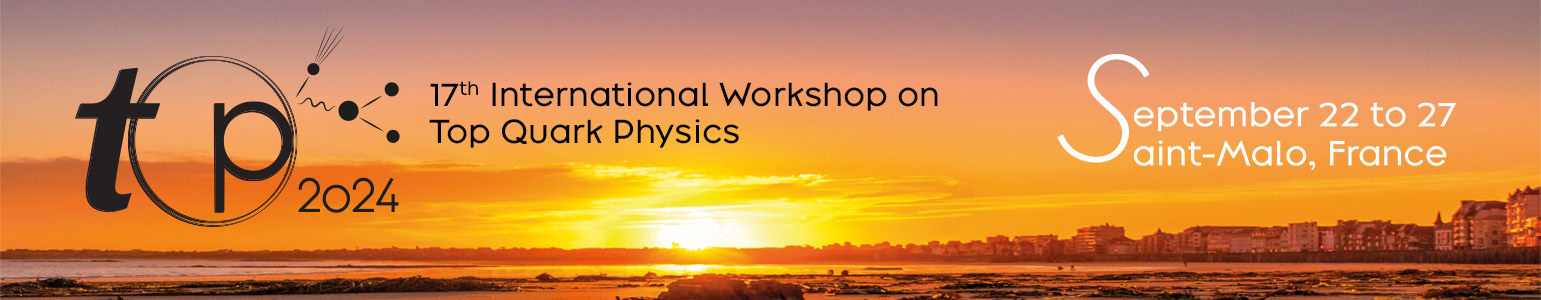}
  \end{minipage}
  &
  \begin{minipage}{0.55\textwidth}
    \begin{center} \hspace{5pt}
    {\it The 17th International Workshop on\\ Top Quark Physics (TOP2024)} \\
    {\it Saint-Malo, France, 22-27 September 2024
    }
    \doi{10.21468/SciPostPhysProc.?}\\
    \end{center}
  \end{minipage}
\end{tabular}
}
\end{center}

\section*{\color{scipostdeepblue}{Abstract}}
\textbf{\boldmath{%
The ATLAS collaboration at the LHC has published inclusive cross-section measurements for the single-top and \ttbar production modes at center-of-mass energies of $\sqrt{s} = 5.02, 8.16$, $13$, and $13.6$ TeV. Single-top measurements are conducted in the $t$-channel and $tW$ channel. In addition to the nominal cross-section measurements, various measurements of other interesting observables such as the $V_{tb}$ element of the Cabibbo Kobayashi Maskawa (CKM) quark-mixing matrix, the ratio of the inclusive cross-sections between $tq$ and $t\overline{q}$, the ratio of inclusive cross-sections between $t\overline{t}$ and $Z\rightarrow \ell\ell$, and the nuclear modification factor (defined as the ratio of the inclusive $t\overline{t}$ cross section in heavy-ion collisions to the inclusive $t\overline{t}$ cross-section in $pp$ collisions) are also reported. These results are compared to their corresponding SM predictions, calculated at (N)NLO in QCD. All results are in good agreement with SM predictions. 
}}

\vspace{\baselineskip}

\noindent\textcolor{white!90!black}{%
\fbox{\parbox{0.975\linewidth}{%
\textcolor{white!40!black}{\begin{tabular}{lr}%
  \begin{minipage}{0.6\textwidth}%
    {\small Copyright attribution to authors. \newline
    This work is a submission to SciPost Phys. Proc. \newline
    License information to appear upon publication. \newline
    Publication information to appear upon publication.}
  \end{minipage} & \begin{minipage}{0.4\textwidth}
    {\small Received Date \newline Accepted Date \newline Published Date}%
  \end{minipage}
\end{tabular}}
}}
}




\section{Introduction}
\label{sec:intro}

The top quark is the heaviest elementary particle of the Standard Model (SM) and is produced in abundance at the CERN Large Hadron Collider (LHC). The top quark carries several unique properties and remains an active area of research at the LHC. Precision measurements of top quark observables serve as stringent tests of the SM, probing the limits of perturbative Quantum Chromodynamics (QCD) at next-to-next leading-order (NNLO) precision. These measurements are then used to constrain various Monte-Carlo (MC) generator parameters towards improving the modeling of various SM backgrounds involving top quarks.

At the LHC, there are two dominant modes of production for top quarks, defined by the number of on-shell top (or antitop) quarks produced. The dominant production mode involves the strong interaction and produces a top-antitop (\ttbar) pair. The next dominant mode of production occurs via the electroweak interaction producing a single on-shell top. Singly-resonant top production can be further categorized into various production modes. The two dominant modes of single-top production are the $t$-channel and $tW$ (top quark produced in association with a $W$ boson) channel. Single-top production is associated with a $t\rightarrow Wb$ vertex which provides the opportunity to measure the \vtb element of the Cabibbo Kobayashi Maskawa (CKM) matrix. The cross-section of single-top production is proportional to the magnitude of the \vtb element.

\begin{figure}[H]
\centering
\includegraphics[width=\linewidth]{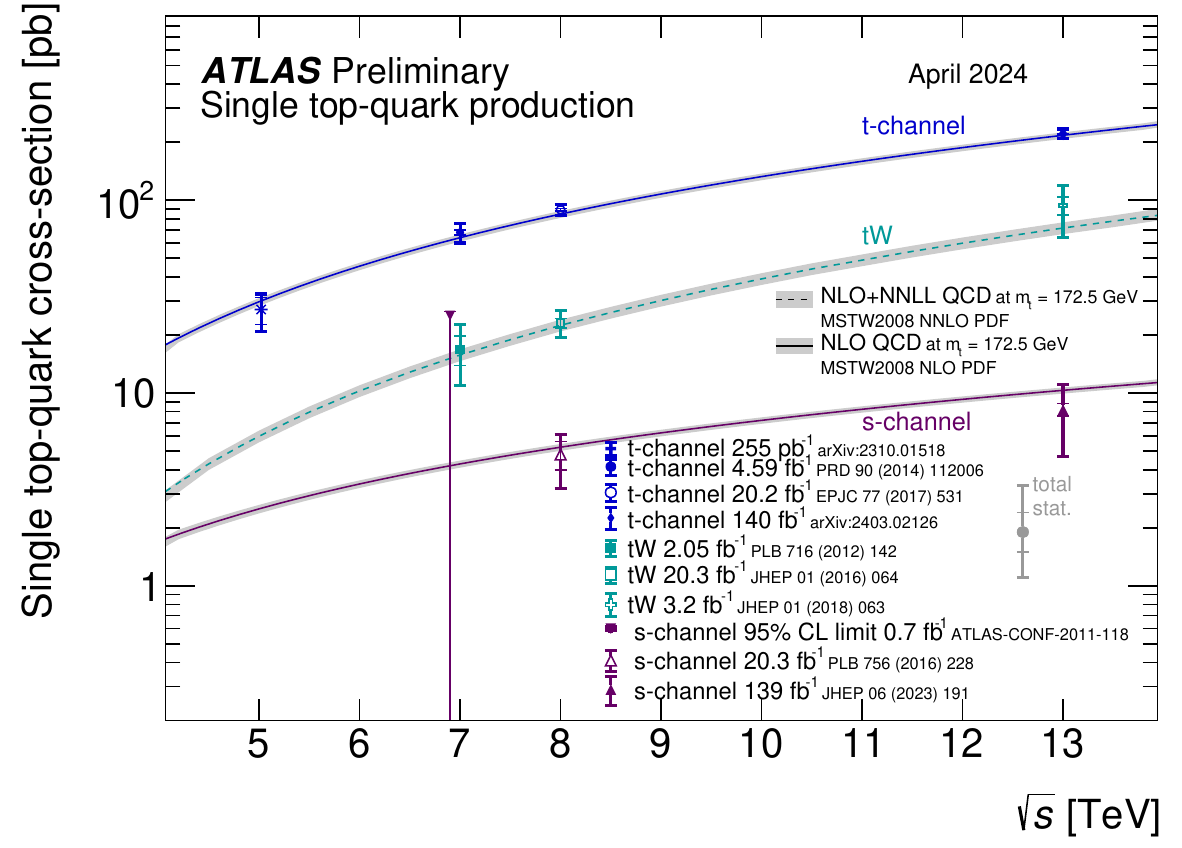}
\caption{The measured single-top cross-sections in ATLAS for various production modes are plotted against the current SM predictions as a function of the center-of-mass energy. The predicted cross-sections are calculated at NLO in QCD. For the $tW$ channel, next-to-next leading log (NNLL) resummation has also been applied. The predictions assume a top quark mass of 172.5 GeV. Figure is taken from public Top Cross-Section Summary plots (April 2024) \cite{ATLASTOPPUB}.}
\label{fig:single_top_xsec}
\end{figure}

\begin{figure}[H]
\centering
\includegraphics[width=\linewidth]{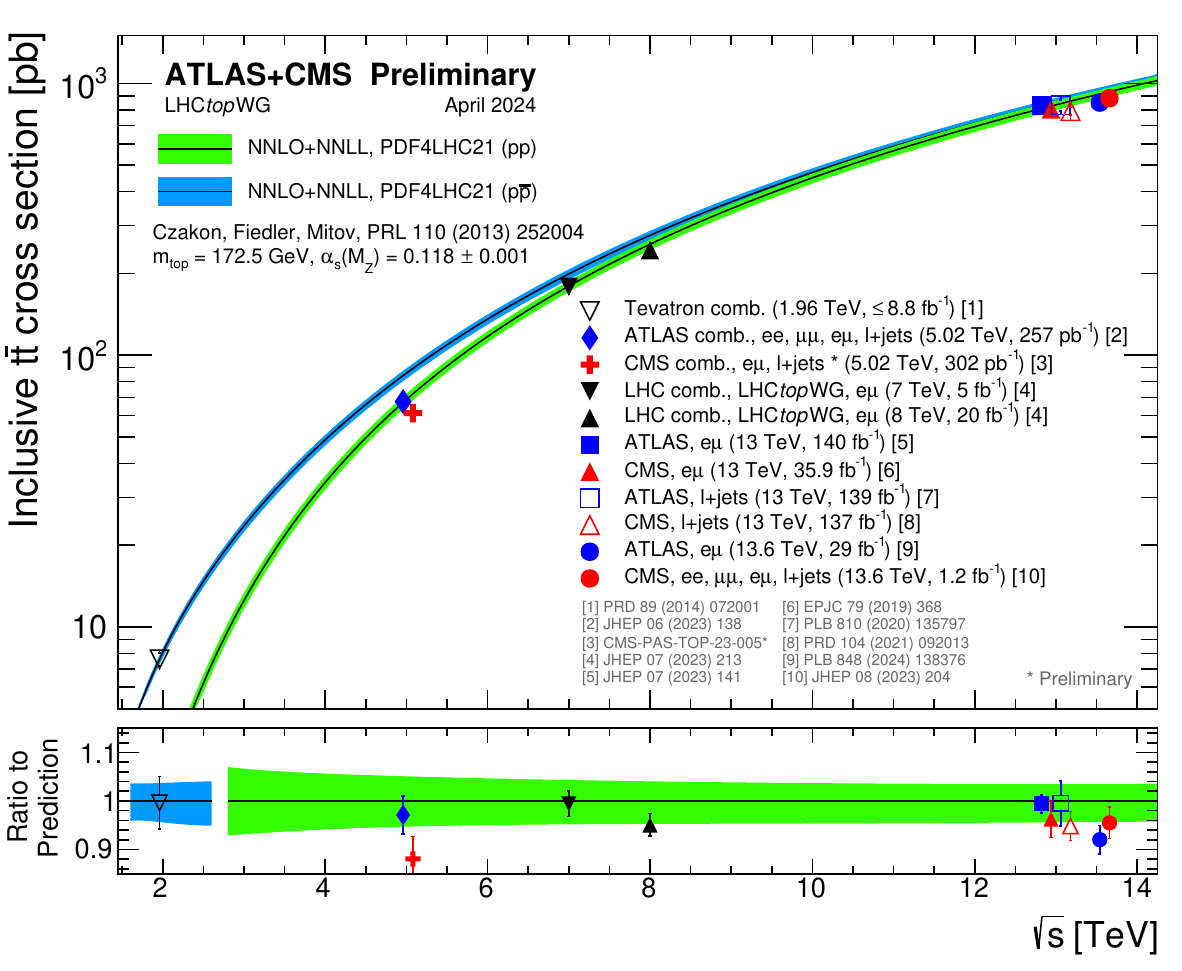}
\caption{The measured \ttbar cross-sections at the LHC and the Tevatron for various \ttbar final states are plotted against the current SM prediction as a function of the center-of-mass energy. The predictions are calculated at NNLO in QCD with NNLL resummation applied. The uncertainty in the predictions come from a combination of renormalisation and factorisation scales, PDFs, and $\alpha_S$. The predictions assume a top quark mass of 172.5 GeV and a strong coupling constant of $\alpha_S = 0.118$. Figure is taken from public Top Cross-Section Summary plots (April 2024) \cite{ATLASTOPPUB}.}
\label{fig:ttbar_xsec}
\end{figure}

This report presents measurements performed by the ATLAS collaboration \cite{ATLAS} of the inclusive cross-sections for the \ttbar and single-top production modes. These measurements are compared to cross-section predictions calculated at NNLO \cite{tChannelNNLO,tWNNLO,ttbarNNLO} as shown in Figure \ref{fig:single_top_xsec} and Figure \ref{fig:ttbar_xsec}.

\section{$tW$ production cross-section measurement at $\sqrt{s} = 13$ TeV}

The inclusive $tW$ production cross-section has been measured at a center-of-mass energy of $\sqrt{s} = 13$ TeV and uses the full Run-2 data set corresponding to an integrated luminosity of $140$ fb$^{-1}$ \cite{tW13TeV}. The analysis is conducted in the dileptonic channel in which the final state comprises two oppositely charged leptons and with opposite flavours ($e^{\pm}\mu^{\mp}$) and at least one $b$-tagged jet (\bjet). The major source of background events come from $t\overline{t}$ events, with minor contributions from $W$+jets, $Z$+jets, and diboson events. After selection, events are organized into three orthogonal signal regions: \textit{1j1b}, \textit{2j1b}, and \textit{2j2b}. The signal regions are defined as \textit{NjMb} where \textit{N} denotes the number of jets, of which \textit{M} are $b$-tagged.

After selection, a Boosted Decision Tree (BDT) is trained to differentiate between signal $tW$ events and background \ttbar events. Distributions of the BDT response variable are constructed in each of the three signal regions. A binned profile likelihood fit is then applied to each distribution to extract the signal yield and compute the cross-section. To relax stringent constraints placed on certain uncertainties (e.g., uncertainty on the double-counting between single-top and events evaluated using the Diagram Removal (DR) and Diagram Subtraction (DS) methods), the fit is applied to a restricted range of the BDT response distribution. This results in a more reliable measurement at the cost of loss of precision. Nevertheless, the post-fit uncertainties are reduced from approximately 10\% to 3-4\% while accounting for all correlations between uncertainties. Systematic uncertainties are included as nuisance parameters in the fit. The dominant sources of uncertainty correspond to \ttbar modeling (13.2\%), Jet Energy Scale (JES) (12.0\%), and \etmiss reconstruction and calibration (11.0\%). The total inclusive cross-section is measured to be 
\begin{equation}
    \sigma_{tW} = 75 \pm 1(\text{stat})\ ^{+15}_{-14} (\text{syst}) \pm 1 (\text{lumi})\ \text{pb}
\end{equation}
This value agrees well with the SM prediction of $\sigma_{tW}^{\text{theo}} = 79.3\ ^{+2.9}_{-2.8}$ pb. The ratio of the measured cross-section to the theoretical cross-section corresponds to a value of \flvvtb$^2$. Here, $f_{LV}$ represents a left-handed form factor that accounts for non-SM contributions. The \flvvtb value is measured to be 
\begin{equation}
    |f_{LV}V_{tb}| = 0.97 \pm 0.10
\end{equation}
This measurement is in agreement with the SM prediction of unity.

\section{$t$-channel production cross-section measurement at $\sqrt{s} = 13$ TeV}

Single-top production via the exchange of a virtual $W$ ($t$-channel) is known as the "golden" channel defined by the largest single-top production cross-section. The ATLAS collaboration has produced a measurement of the inclusive single-top and single-antitop cross-sections in this channel at a center-of-mass energy of $\sqrt{s} = 13$ TeV using 140 fb$^{-1}$ of ATLAS data \cite{tChannel13TeV}. Events are required to contain exactly one lepton, exactly two jets (with exactly one \bjet), and additional cuts on \etmiss and \mtw to reduce multijet backgrounds. The $W$ transverse mass (\mtw) is defined as

\begin{equation}
    m_{\text{T}}(W) = \sqrt{2p_{\mathrm{T}}(\ell)E_{\mathrm{T}}^{\mathrm{miss}}(1-\cos\Delta\phi(\Vec{p}_{\mathrm{T}}^{\mathrm{miss}}, \ell))}
\end{equation}
The $W$ transverse mass is calculated using the transverse momentum of the charged lepton ($p_{\mathrm{T}}(\ell)$), the missing transverse momentum ($\Vec{p}_{\mathrm{T}}^{\mathrm{miss}}$), and the difference of the azimuthal angles of the $\Vec{p}_{\mathrm{T}}^{\mathrm{miss}}$ vector and the charged lepton vector ($\Delta\phi(\Vec{p}_{\mathrm{T}}^{\mathrm{miss}}, \ell)$). The event selection defines two signal regions depending on the sign of the lepton: \textit{SR plus} and \textit{SR minus}.

A neural network, trained on a combination of $tq$ and $t\overline{q}$ events, is used to separate signal and background events in the two signal regions. A discriminatory variable ($D_{nn}$) is formed by combining several kinematic observables. Distributions of $D_{nn}$ in the two signal regions are then entered into a binned profile likelihood fit to extract the cross-section measurements for $tq$ ($\sigma(tq)$) and $t\overline{q}$ ($\sigma(t\overline{q})$). The corresponding measured cross-sections are as follows.
\begin{align}
    \sigma(tq) &= 137\pm 8\ \text{pb}\\
    \sigma(t\overline{q}) &= 84^{+6}_{-5}\ \text{pb}
\end{align}
The measured cross-sections are in good agreement (within uncertainty) with the following SM predictions at 13 TeV.
\begin{align}
    \sigma(tq)^{\text{theo}} &= 134.2^{+2.6}_{-1.7}\ \text{pb}\\
    \sigma(t\overline{q})^{\text{theo}} &= 80.0^{+6}_{-5}\ \text{pb}
\end{align}
It is also useful to report a measurement of the ratio between $\sigma(tq)$ and $\sigma(t\overline{q})$, $R_{t} = \sigma(tq) / \sigma(t\overline{q})$. In doing so, several common uncertainties cancel out resulting in a very precise measurement. This ratio can then be used in future various Parton Distribution Functions (PDF) if the measured value is included in future fits. The measured $R_t$ is 
\begin{equation}
    R_{t} = 1.636^{+0.036}_{-0.034}
\end{equation}
Finally, an independent measurement of \flvvtb is also reported which is again in agreement with the SM prediction of unity
\begin{equation}
    |f_{LV}V_{tb}| = 1.015 \pm 0.031
\end{equation}

\section{$t$-channel production cross-section measurement at $\sqrt{s} = 5.02$ TeV}

The single-top and single-antitop cross-sections are measured at a special center-of-mass energy of 5.02 TeV using 255 pb$^{-1}$ of ATLAS data \cite{tChannel502}. This data sample was collected in a low pileup environment where the mean number of inelastic $pp$ collisions was approximately $2$. This provides an independent test of the SM in an environment with different backgrounds and detector uncertainties. Events are selected if they contain exactly one lepton, exactly two jets (with exactly one \bjet), and additional cuts on \etmiss and \mtw to reduce multijet backgrounds. Two signal regions are defined depending on the sign of the lepton: $\ell^+$\textit{+ jets} and $\ell^-$\textit{+ jets}.

A BDT is trained to separate signal and background events. A profile likelihood fit is applied to the distribution of the BDT response variable in each of the two signal regions. Uncertainties are included in the fit as additional nuisance parameters. These uncertainties are either estimated directly using 5.02 TeV sample or extrapolated from a high pileup 13 TeV sample. The resulting dominant uncertainties are statistical (16\%), single-top modeling (8.6\%), and misidentified leptons (6.3\%). The resulting cross-sections extracted from the fit are as follows.
\begin{align}
    \sigma(tq) &= 19.8^{+3.9}_{-3.1}(\text{stat})\ ^{+2.9}_{-2.2}(\text{syst})\ \text{pb}\\
    \sigma(t\overline{q}) &= 7.3^{+3.2}_{-2.1}(\text{stat})\ ^{+2.8}_{-1.5}(\text{syst})\ \text{pb}
\end{align}
These cross-sections are in agreement with SM predictions but come with much larger uncertainties than similar measurements at higher center-of-mass energies. Again, measurements of $R_t$ and \flvvtb are also reported. 
\begin{align}
    R_t &= 2.73^{+1.43}_{-0.82}(\text{stat})\ ^{+1.01}_{-0.29}(\text{syst})\\
    |f_{LV}V_{tb}| &= 0.94^{+0.11}_{-0.10}
\end{align}

\section{\ttbar production cross-section measurement at $\sqrt{s} = 13.6$ TeV}

\ttbar production cross-section has been measured using relatively recent Run 3 data collected in 2022 at 13.6 TeV and 29 fb$^{-1}$ of ATLAS data \cite{ttbarRun3}. This analysis measures both the inclusive \ttbar ($\sigma_{t\overline{t}}$) and $Z$-boson ($\sigma_{Z\rightarrow \ell\ell}$) production cross-sections and the ratio between the two ($R_{t\overline{t}/Z}$). In analogy with the ratio defined for the single-top case, $R_{t\overline{t}/Z}$ benefits from the cancellation of several common uncertainties and produces a very precise measurement. This ratio also has a particular sensitivity to the gluon-to-quark PDF ratio.

Event selection for top quark pair production requires two oppositely flavored leptons ($e\mu$) and exactly one or two \bjets. For $Z$-boson decays, two same flavored leptons ($ee$ or $\mu\mu$) are required with the additional requirement that the invariant mass ($m_{\ell\ell}$) of the lepton pair must falls within $66 < m_{\ell\ell} < 116$ GeV. This selection yields a nearly 100\% pure sample of $Z$-boson decay events and allows an estimate of the $Z$-boson contribution. A profile likelihood fit is then applied to the distribution of the event yields in each of the four signal regions. The  measured cross-sections and ratio extracted from the fit are
\begin{align}
    \sigma_{t\overline{t}} &= 850 \pm 3(\text{stat}) \pm 18(\text{syst}) \pm 20(\text{lumi})\ \text{pb}\\
    \sigma_{Z\rightarrow \ell\ell} &= 744 \pm 11(\text{stat}+\text{syst}) \pm 16(\text{lumi})\ \text{pb}\\
    R_{t\overline{t}/Z} &= 1.145 \pm 0.003(\text{stat}) \pm 0.021(\text{syst}) \pm 0.002(\text{lumi})
\end{align}
The dominant sources of uncertainty correspond to luminosity (2.3\%) and lepton reconstruction (1.4\%). These measurements agree with the SM predictions within one standard deviation. 
\begin{align}
    \sigma_{t\overline{t}}^{\text{theo}} &= 924^{+32}_{-40}(\text{scale+PDF+}\alpha_S)\ \text{pb}\\
    \sigma_{Z\rightarrow \ell\ell}^{\text{theo}} &= 746^{+21}_{-22}(\text{scale+PDF+}\alpha_S)\ \text{pb}\\
    R_{t\overline{t}/Z}^{\text{theo}} &= 1.238^{+0.063}_{-0.071}(\text{scale+PDF+}\alpha_S)
\end{align}

\section{\ttbar production cross-section measurement in proton-lead collisions at $\sqrt{s_{NN}} = 8.16$ TeV}

The top quark pair production cross-section has been measured in proton-lead collisions at a nucleon-nucleon center-of-mass energy of $\sqrt{s_{NN}} = 8.16$ TeV and an integrated luminosity of 165 nb$^{-1}$ \cite{ttbarPPb}. In $p+Pb$ collisions, top quark measurements probe kinematic regions of nuclear PDFs that are not readily accessed by other measurements due to limited statistics. This analysis defines both dileptonic and semileptonic signal regions. The dilepton signal regions ($2\ell1b$, $2\ell2b$) are defined by two oppositely charged leptons. Events are discarded if the invariant mass of the two leptons fall within $80 < m_{\ell\ell} < 100$ GeV to remove $Z$-boson contributions. There are additional requirements of at least two jets (of which there must be at least one \bjet). The semilteptonic signal regions ($1\ell1b$ \textit{e+jets}, $1\ell2bincl$ \textit{e+jets}, $1\ell1b\ \mu$\textit{+jets}, $1\ell2bincl\ \mu$\textit{+jets}) require exactly one lepton and at least four jets (with at least one \bjet). 

The main backgrounds in the dilepton signal regions include $Z$+jets and single-top $tW$ processes while the main background in the semileptonic regions corresponds to non-prompt and misidentified leptons (estimated using the matrix-method). The expected yields of the signal are extracted from a profile likelihood fit to a $H_{T}^{\ell,j}$ (scalar sum of the transverse momentum of the leptons and jets) observable in the six signal regions. The measured inclusive \ttbar cross-section is 
\begin{equation}
    \sigma_{t\overline{t}} = 58.1 \pm 2.0(\text{stat})\ ^{+4.8}_{-4.4}(\text{syst})\ \text{nb}
\end{equation}
The measured cross-section is compared to predictions provided by various PDF sets. The dominant uncertainties in the measurement arise from JES (4.6\%) and \ttbar generator (4.5\%) uncertainties. A measurement of the nuclear modification factor is also reported. This factor is defined as the ratio of the total \ttbar cross-section in $p+Pb$ ($\sigma_{t\overline{t}}^{p+Pb}$) to the cross-section in nominal $pp$ ($\sigma_{t\overline{t}}^{pp}$) collisions at $\sqrt{s} = 8$ TeV normalized by the mass number of lead ($A_{Pb} = 208$). The $pp$ cross-section at $8$ TeV is extrapolated to the center-of-mass energy of the $p+Pb$ system. 
\begin{equation}
    R_{pA} = \frac{\sigma_{t\overline{t}}^{p+Pb}}{A_{Pb}\cdot\sigma_{t\overline{t}}^{pp}}
\end{equation}
The measured nuclear modification factor is in agreement with the SM prediction of unity.
\begin{equation}
    R_{pA} = 1.090 \pm 0.039(\text{stat})\ ^{+0.094}_{-0.087}(\text{syst})
\end{equation}

\section{Conclusion}

Proton-proton collisions at the LHC produce top quarks in abundance. Cross-section measurements for various top processes have been conducted by the ATLAS collaboration using Run 2 and Run 3 data. These measurements allow for a direct comparison with SM predictions at the NNLO level. Cross-sections were measured for both the single-top and \ttbar production modes. Measurements of single-top production cross-sections were conducted for both the $t$-channel and $tW$ processes at the nominal center-of-mass energies of $\sqrt{s} = 13$ TeV. The final cross-sections are measured with relative uncertainties of 20\% ($t$-channel) and 7\% ($tW$). An independent measurement of $t$-channel production has also been measured at a center-of-mass energy of $\sqrt{s} = 5.02$ TeV. The measured cross-section is consistent with SM predictions but is accompanied by a much higher relative uncertainty of 58\%. Each of the single-top analyses has also reported independent measurements of the \vtb CKM matrix element.

Measurements of \ttbar production remain one of the most precise measurements due to an abundance of viable top quark pair events. The \ttbar production cross-section has been measured using Run 3 data collected in 2022. As expected the final cross-section is accompanied by a very low relative uncertainty of 3\%. \ttbar production has also been observed and measured in $p+Pb$ collisions. This measurement comes with a relative uncertainty of 9\%. 

All of the results presented in this report are consistent with SM predictions. 

\bibliography{SciPost_Example_BiBTeX_File.bib}


\end{document}